\documentclass[ prl,twocolumn,floatfix,amsmath,amssymb,aps,showpacs,longbibliography,superscriptaddress]{revtex4-1}
\usepackage{graphicx}
\usepackage{xcolor}
\usepackage{siunitx}
\usepackage{hyperref}
\usepackage[normalem]{ulem}
\usepackage{here}
\usepackage {amsmath}
\usepackage{mathrsfs}
\usepackage{bm}
\usepackage{color} \usepackage{graphicx} \graphicspath{{Figures/}}
\usepackage{comment} \usepackage{color}
\usepackage[normalem]{ulem}
\renewcommand{\epsilon}{\varepsilon}

\newcommand{\tn}[1]{\textnormal{#1}} 

\begin{document}

\title{Active motility and wetting cooperatively regulate liquid--liquid phase separation}

\author{Dixi Yang}
\thanks{These authors contributed equally}
\affiliation{Advanced Materials Thrust, The Hong Kong University of Science and Technology (Guangzhou), Guangzhou 511453, China}

\author{Anheng Wang}
\thanks{These authors contributed equally}
\affiliation{Institute of Chinese Medical Sciences \& State Key Laboratory of Mechanism
and Quality of Chinese Medicine, University of Macau, Macau SAR 999078, China}
\affiliation{Department of Pharmacy and Pharmaceutical Sciences, Faculty of Medicine, University of Macau, Macau SAR 999078, China}
\affiliation{Zhuhai UM Science and Technology Research Institute, University of Macau, Hengqin 519031, Guangdong, China}
\affiliation{University of Macau Advanced Research Institute in Hengqin, Hengqin 519031, Guangdong, China}

\author{Chunming Wang}
\email{cmwang@um.edu.mo}
\affiliation{Institute of Chinese Medical Sciences \& State Key Laboratory of Mechanism
and Quality of Chinese Medicine, University of Macau, Macau SAR 999078, China}
\affiliation{Department of Pharmacy and Pharmaceutical Sciences, Faculty of Medicine, University of Macau, Macau SAR 999078, China}
\affiliation{Zhuhai UM Science and Technology Research Institute, University of Macau, Hengqin 519031, Guangdong, China}
\affiliation{University of Macau Advanced Research Institute in Hengqin, Hengqin 519031, Guangdong, China}

\author{Hajime Tanaka}
\email{tanaka@iis.u-tokyo.ac.jp}
\affiliation{Research Center for Advanced Science and Technology, The University of Tokyo, 4-6-1 Komaba, Meguro-ku, Tokyo 153-8904, Japan}
\affiliation{Department of Fundamental Engineering, Institute of Industrial Science, The University of Tokyo, 4-6-1 Komaba, Meguro-ku, Tokyo 153-8505, Japan}

\author{Jiaxing Yuan}
\email{jiaxingyuan@hkust-gz.edu.cn}
\affiliation{Advanced Materials Thrust, The Hong Kong University of Science and Technology (Guangzhou), Guangzhou 511453, China}
\affiliation{The Hong Kong University of Science and Technology, Clear Water Bay, Hong Kong SAR, China}

\date{October 18, 2026}

\begin{abstract}
Liquid--liquid phase separation in aqueous two-phase systems is widespread in physical and biological settings, yet how motile agents regulate its morphology and kinetics remains poorly understood. Combining experiments on \textit{Pseudomonas aeruginosa} in a dextran--polyethylene glycol mixture with hydrodynamic simulations, we show that interfacial wetting couples bacterial motility to phase-separation dynamics. At high dextran fraction, bacteria-driven rotational flows shear inter-droplet gaps and hinder liquid-bridge formation, thereby suppressing coalescence and coarsening. As the dextran fraction decreases, wetting-driven bridging becomes increasingly important, producing an intermediate chain-like state and, ultimately, branched bacterial clusters in which activity instead accelerates capture and aggregation. A biofilm-associated protein provides an independent wetting component that similarly promotes bacterial clustering. These results reveal a composition-controlled crossover between activity-induced suppression of coalescence and wetting-dominated bridging.
\end{abstract}

\maketitle

\noindent
{\it \bf Introduction.---} Phase separation, in which a homogeneous mixture spontaneously segregates into distinct phases, plays a central role in soft matter~\cite{doi2013soft,tanaka1993,Bailey2007,tanaka2012viscoelastic,tateno2021,Yuan2023Mechanical,chen2023charge}, biological cells~\cite{Brangwynne2009,li2012phase,Simon2019,Sarah2019,tanaka2022Viscoelastic}, and industrial materials~\cite{Tanaka2009,Gallegos1999,wang2019progress}. Liquid--liquid phase separation in aqueous two-phase systems (ATPSs) has attracted particular attention because of its relevance to intracellular compartmentalization~\cite{Brangwynne2009,hyman2014liquid} and biotechnology~\cite{teixeira2018emerging,ma2020cell,zhang2025aqueous}. After phase separation, interfacial tension generally drives domain coarsening toward structures with reduced interfacial area~\cite{onuki2002phase}.

While phase separation in passive ATPS has been extensively explored, recent work has focused on the living control of domain patterns via the introduction of active elements~\cite{Stenhammar2015,Smrek2017,zhang2021active,arnold2023active,Bhattacharyya2023,yuan2024colloid,jambon2024phase,sun2025engineering}, such as self-propelled bacteria~\cite{grober2023unconventional,Ning2023,cheon2024partitioning,black2025capillary}. These motile agents generate hydrodynamic interactions (HIs) as they swim through the fluid~\cite{drescher2011fluid,yeomans2014introduction,hu2015modelling,shen2019hydrodynamic,shaebani2020computational}, inducing long-range couplings. It is well established that bacteria can exhibit collective flowing behaviors~\cite{wu2017transition,xu2019self,li2024robust} and self-organization~\cite{furukawa2014activity,chen2015dynamic,theers2018clustering} governed by such HIs. Previous studies have highlighted that HIs play a crucial role in phase separation dynamics~\cite{tanaka1997viscoelastic,tanaka2017phase,arnold2023active}.
 
In addition to HIs, preferential wetting and interfacial partitioning can strongly influence phase-separation dynamics. Previous studies have established how passive colloids partition between coexisting liquids and modify coarsening~\cite{dobbs1992capillary,bonn2009wetting,thijssen2010demixing,zhang2024wetting,tanaka2001interplay,araki2008dynamic,araki2006wetting}. More recently, Jeong \textit{et al.} showed that bacterial motility modifies partitioning in a strongly confined DEX--PEG ATPS in which droplet coarsening is suppressed by geometry~\cite{cheon2024partitioning}. What remains unclear is how, in an unconstrained phase-separating fluid, wetting and motility couple dynamically to control liquid-bridge formation, coalescence, and morphology selection.

In this Letter, we combine three-dimensional experiments on \textit{Pseudomonas aeruginosa} (\textit{P. aeruginosa}) with fluid particle dynamics (FPD) simulations~\cite{tanaka2000,Furukawa2018} to determine how interfacial wetting and self-propulsion jointly regulate morphology and coarsening kinetics in a DEX--PEG ATPS. By varying the DEX volume fraction, we uncover a sequence of self-spinning droplets, elongated droplet chains, and wetting-bridged branched bacterial clusters. We interpret this sequence as a composition-controlled shift in the competition between wetting-driven liquid-bridge formation and its disruption by activity-induced rotational flow. In the droplet regime, interfacially localized bacteria drive droplet rotation; the resulting tangential flow shears the inter-droplet gap, hinders liquid-bridge formation, and thereby suppresses coalescence and slows coarsening. At intermediate DEX fraction, wetting-mediated bridge formation becomes increasingly important while the rotational flow still opposes complete coalescence; this competition is consistent with the formation of long-lived, connected chain-like structures. When DEX becomes strongly limited, wetting-mediated bridging dominates and the scarce DEX-rich phase connects neighboring bacteria, producing branched clusters. Chemical-potential gradients and the associated flow provide a transient transport mechanism for bringing the wetting phase into the inter-bacterial region, whereas the bound state is stabilized by wetting thermodynamics after these gradients relax. To demonstrate the broader relevance of this mechanism, we further show that a biofilm-associated protein can act as an interfacial ``wetting glue,'' promoting bacterial clustering even in dilute suspensions. Together with our companion paper~\cite{yang2026b}, which establishes how interfacial localization converts bacterial motility into coherent droplet rotation, the present results show how active interfacial flow competes with wetting-driven bridge formation to regulate morphology and kinetics during LLPS.

\noindent
{\it \bf Experimental.---}
To quantify the wetting properties of \textit{P. aeruginosa}, we measure contact angles by depositing a DEX droplet onto the PEG--bacteria interface (see {\bf End Matter} and Fig.~S1 in \cite{Supple} for experimental details). The contact angle, $\theta \approx 50^\circ$, indicates that \textit{P. aeruginosa} are amphiphilic, with affinity for both PEG and DEX phases and a slight preference for DEX (Fig.~\ref{fig:figureA1}(a)). Control measurements using water, DEX, and PEG droplets on the air--bacteria interface show the strongest affinity for water. This shows that the three-phase contact angle measured within the ATPS is the relevant metric for predicting bacterial partitioning between DEX and PEG.

\begin{figure}[t!]
\centering \includegraphics[width=8.5cm]{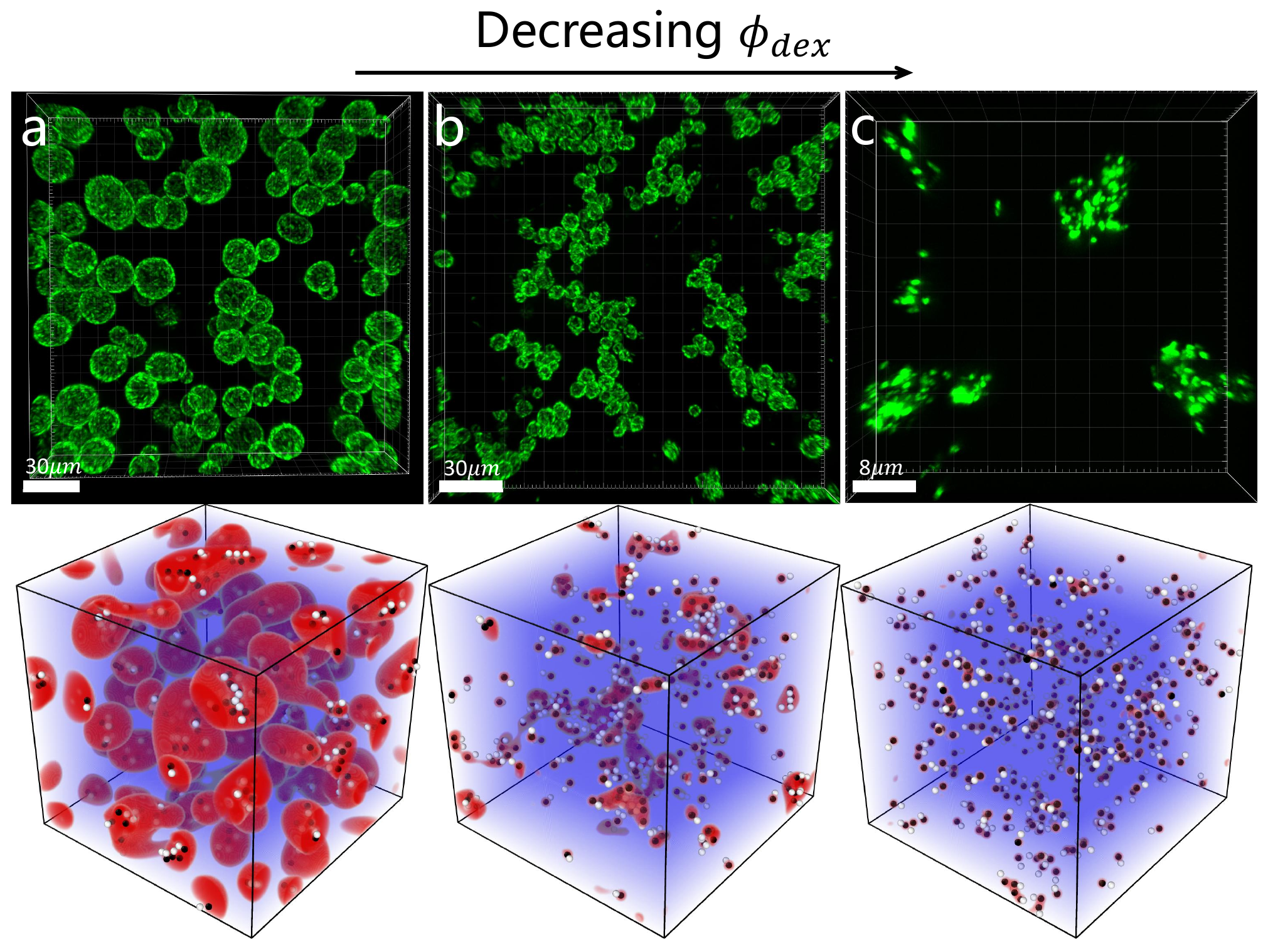}
\caption{Morphology transition in an ATPS with \textit{P. aeruginosa} ($\phi_{\mathrm{bac}}=0.01$) upon decreasing DEX volume fraction~$\phi_\text{dex}$.
(a) Self-spinning droplets ($\phi_\text{dex}=0.05$).
(b) Droplet chains ($\phi_\text{dex}\approx\phi_{\mathrm{bac}}$).
(c) Wetting-bridged branched bacterial clusters ($\phi_\text{dex}=0.001$).
Top: experimental images of GFP-tagged bacteria.
Bottom: corresponding simulation snapshots (black: head; white: tail; red: DEX; blue: PEG).
}
\label{fig:figure1}
\end{figure}

\begin{figure}[t!]
\centering \includegraphics[width=8.5cm]{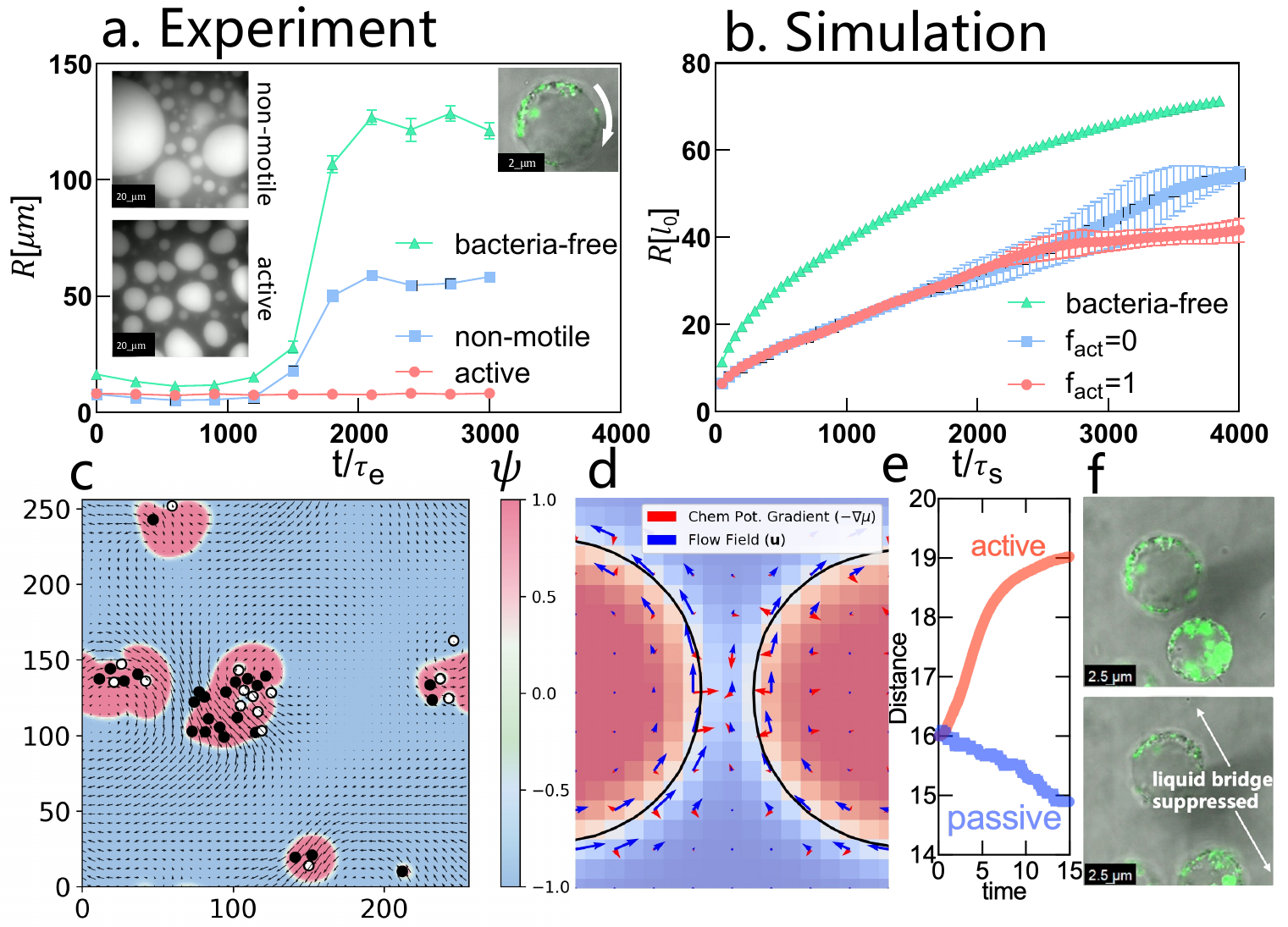}
\caption{Dynamics and coarsening mechanism of self-spinning droplets.
(a) Experimental evolution of the average droplet radius $R$ (time scaled by $\tau_\text{e}=\ell_\text{bac}/v_\text{bac}\approx0.1\,$s, with $\ell_\text{bac}\approx1.5\,\mu$m and $v_\text{bac}\approx15\,\mu$m/s) for $\phi_\text{dex}=0.05$, $\phi_\text{bac}=0.01$, and a bacteria-free control.
Inset: comparison of droplet radius for active and non-motile systems at $t=2400\tau_\text{e}$ (left) and a rotating droplet (right).
(b) Simulated average droplet radius $R$, showing slower coarsening in the active case.
(c) Composition field $\psi$ and velocity field $\bm{v}$, showing droplet rotation.
(d) Physical origin of arrested coalescence by interfacial flows. The rotational flow is predominantly transverse to the chemical-potential-driven transport in the gap and continuously shears and rearranges the inter-droplet region, thereby hindering the establishment of a DEX-rich liquid bridge. (e) Time evolution of the center-to-center distance between the two deformable droplets. The red curve corresponds to the system with active rotation, averaged over 1000 random orientations, whereas the blue curve denotes the thermodynamically driven passive control. (f) Time-lapse experimental images of nearby self-spinning droplets without forming stable liquid bridges. The curves in (a) and (b) are averaged over 3 independent samples and are shown as the mean $\pm$ standard deviation; where the error bars are not visible, they are smaller than symbols.
}
\label{fig:figure2}
\end{figure}

We next perform experimental observations using a spinning-disk confocal microscope (Fig.~\ref{fig:figureA1}(b)), enabling real-time \emph{three-dimensional} visualization of phase separation patterns (see \cite{Supple} for details). Even a small amount of \textit{P. aeruginosa} markedly alters the phase-separation dynamics compared with the bacteria-free ATPS, which exhibits passive droplet coarsening without autonomous rotation ({\bf Movie~S1}). At a low bacterial volume fraction $\phi_\text{bac}=0.01$, decreasing the DEX volume fraction~$\phi_\text{dex}$ induces a sequence of dynamic morphologies, including self-spinning droplets (Fig.~\ref{fig:figure1}(a), top; {\bf Movie~S2}), elongated droplet chains (Fig.~\ref{fig:figure1}(b), top), and wetting-bridged branched bacterial clusters (Fig.~\ref{fig:figure1}(c), top).

We observe the bacteria preferentially localize at the interfacial region (Fig.~\ref{fig:figureA2}), rather than in the bulk of droplet~\cite{cheon2024partitioning}, confirming their amphiphilic nature. The resulting non-equilibrium patterns---from self-spinning droplets ({\bf Movie~S3--S5}) to droplet chains ({\bf Movie~S6}) and branched bacterial clusters ({\bf Movie~S7})---remain robust even at higher bacterial volume fractions $\phi_\text{bac}=0.05$--$0.1$. These results establish a robust route for the living control of domain morphology and reveal a highly dynamic, composition-dependent morphology transition not previously reported.

\noindent
{\it \bf Hybrid particle--field simulations.---} 
To gain mechanistic insight, we develop a coarse-grained particle--field hybrid model for active bacteria in a binary fluid mixture. Each bacterium is modeled as a pusher-type swimmer composed of a head, a tail, and a virtual particle arranged in a rod-like geometry~\cite{furukawa2014activity}, with activity introduced via a force dipole of magnitude $f_{\text{act}}$ acting between the tail and the virtual particle. The binary mixture is described by a Ginzburg--Landau free-energy functional in terms of a compositional order parameter $\psi$, where $\psi=1$ and $-1$ correspond to DEX- and PEG-rich phases, respectively. The evolution of $\psi$ obeys a conservative advection--diffusion equation coupled to the Navier--Stokes equation, solved using the FPD method~\cite{tanaka2000,Furukawa2018,Harlow1965}. Bacteria--fluid wetting is incorporated through surface-coupling terms with affinity coefficients $\gamma_1$ and $\gamma_2$ for the head and tail. Consistent with the observed interfacial localization (Fig.~\ref{fig:figureA2}) and slight preference for the DEX phase (Fig.~S1 in \cite{Supple}), we set $\gamma_1=-4$ and $\gamma_2=2$.

As the average composition $\bar\psi$ decreases from $-0.6$ to $-0.9$ and $-0.98$, the simulations capture the same broad progression from droplet-like to increasingly connected wetting-dominated structures (Fig.~\ref{fig:figure1}, bottom). The droplet and branched-cluster regimes are reproduced clearly, whereas the experimental chain morphology is captured only qualitatively within the accessible particle number and scale separation. Taken together, the simulations suggest a composition-dependent shift in the balance between activity-induced hydrodynamic flow and preferential-wetting-driven bridging.

\noindent
{\it \bf Active rotational flows suppress droplet coalescence.---}
In the droplet-forming regime ($\phi_{\text{dex}} > \phi_{\text{bac}}$; Fig.~\ref{fig:figure1}a), isolated DEX-rich droplets exhibit spontaneous rotation, signaling bacterial activity. In a companion paper~\cite{yang2026b}, we show that this rotation arises from bacteria localized at the interface: interfacial wetting constrains the swimmers and biases their motion, producing polar ordering that rectifies self-propulsion into a coherent torque. Unlike bulk-driven rotational droplets that require high bacterial densities~\cite{ramos2020bacteria,rajabi2021directional}, our system operates via an interfacial mechanism in which a sparse population of bacteria generates the torque responsible for sustained rotation.

Beyond the specific case of self-spinning droplets~\cite{yang2026b}, here we focus on how activity regulates domain coarsening dynamics. Experimentally, we quantify this effect by tracking the droplet radius~$R$ (see \cite{Supple} for details). We find that activity suppresses coarsening, yielding significantly smaller droplets than in systems with non-motile bacteria (Fig.~\ref{fig:figure2}(a)). This kinetic frustration is evident in the inset of Fig.~\ref{fig:figure2}(a) and is further supported by the delayed macroscopic stratification observed with active \textit{P. aeruginosa} (Fig.~\ref{fig:figureA3}) where the system fully separates into a PEG-rich upper phase and a DEX-rich lower phase under gravity.

The simulated average droplet radius~$R$ (Fig.~\ref{fig:figure2}(b); see \cite{Supple} for details) likewise shows slower late-stage growth in the active system, consistent with the experimental trend. We use the simulation as a minimal mechanistic model rather than as a quantitative fit to the experimental kinetics. The simulations do not include transient macroscopic flows generated during sample preparation or buoyancy-driven convection arising from the DEX--PEG density mismatch. Residual flow after mixing may therefore contribute to the early-time plateau in Fig.~\ref{fig:figure2}(a). We consequently focus on the robust late-stage trend and on how interfacial activity suppresses liquid-bridge formation and droplet coalescence.

What is the origin of this slowed coarsening? To isolate the role of rotational interfacial flow, we simulate two deformable DEX-rich droplets and compare otherwise identical cases with and without active interfacial rotation (Fig.~\ref{fig:figure2}(d)--(e) and Fig.~\ref{fig:figureA4}). Without rotation, the intervening PEG-rich layer drains, a DEX-rich bridge forms, and the droplets coalesce (Fig.~\ref{fig:figureA4}(a)). With rotation, tangential flow continuously shears and rearranges the inter-droplet gap, hindering the establishment and growth of the liquid bridge and thereby arresting coalescence (Fig.~\ref{fig:figureA4}(b)). The flow in the gap is predominantly transverse to the chemical-potential-driven transport (Fig.~\ref{fig:figure2}(d)), so that active advection persistently reorganizes the region in which radial composition transport must build the bridge. Correspondingly, the center-to-center distance remains increasing in the rotating case, whereas the passive control readily coalesces (Fig.~\ref{fig:figure2}(e)). For the counter-rotating configuration examined here, the interfacial flows reinforce one another within the gap, producing strong advective rearrangement and hindering liquid-bridge formation (Fig.~\ref{fig:figureA4}(d)). Thus, the mechanism of the slowed coarsening is activity-induced disruption of liquid-bridge formation between deformable, phase-separating droplets.

Consistent with this picture, time-lapse imaging in experiments shows that adjacent rotating droplets fail to form a stable liquid bridge and subsequently move apart under active interfacial flow (Fig.~\ref{fig:figure2}(f); {\bf Movie~S8}). We note that such effective separation is not expected for ideal rigid rotors undergoing Stokes rotation, whose leading rotlet flow is tangential. In the present system, however, the rotating interfaces are deformable and dynamically coupled to composition transport. These additional degrees of freedom  enable effective radial separation even in the Stokes regime while simultaneously hindering the liquid-bridge formation required for coalescence.

Coalescence events are much less frequent in the available movies of the active emulsion ({\bf Movie~S9}) than in the non-motile ({\bf Movie~S10}) and bacteria-free ({\bf Movie~S11}) controls. Droplet-number statistics quantitatively support the observation: over 12 minutes, the motile system shows the smallest reduction in droplet number, the non-motile system an intermediate reduction, and the bacteria-free control the largest (Fig.~S2 in \cite{Supple}). Relative to the bacteria-free system, passive interfacial adsorption of non-motile bacteria already slows coarsening, consistent with a surfactant-like stabilization (Fig.~\ref{fig:figure2}(a)), whereas motility provides an additional dynamic stabilization by disrupting liquid-bridge formation (Fig.~\ref{fig:figure2}(d)--(e) and Fig.~\ref{fig:figureA4}). Thus, the suppression of coalescence increases in the order bacteria-free $<$ non-motile $<$ active, revealing cooperative contributions from passive interfacial adsorption and active flows. These findings connect microscopic interfacial flows to macroscopic phase-separation kinetics.

{\it \bf Competition regime and droplet-chain morphology.---} As the DEX volume fraction approaches the bacterial volume fraction ($\phi_{\text{dex}} \approx \phi_{\text{bac}}$), isolated rotating droplets give way to chain-like assemblies of interconnected DEX-rich droplets with a characteristic size of $\sim 10\ \mu\text{m}$ (Fig.~\ref{fig:figure1}(b) and {\bf Movie~S12}). The chains are clearly observed experimentally, whereas the simulations generate qualitatively related connected structures but do not quantitatively reproduce the experimental morphology within the accessible particle number and scale separation. We therefore interpret the chain state as an intermediate crossover in which wetting-driven bridge formation is strong enough to establish persistent contacts, while activity-induced rotational flow still hinders complete bridge growth and coalescence. This competition provides a natural route to connected but incompletely merged structures. Under otherwise comparable conditions, such chains are not observed in the non-motile control ({\bf Movie~S13}), supporting the conclusion that preferential wetting alone is insufficient to stabilize the chain state.

The characteristic droplet size in the chain regime can be rationalized with a scaling estimate comparing capillary and active interfacial stresses. The capillary scale is $\Sigma_{\gamma}\sim\gamma/R$, whereas the active shear generated by interfacially confined bacteria may be written $\Sigma_{\mathrm{act}}\sim\chi n_s f_\tn{act}$, where $n_s$ is the areal bacterial density and $0<\chi\lesssim1$ is the fraction of propulsion transmitted coherently into tangential interfacial stress. Their ratio defines an active capillary number, $\mathrm{Ca}_{\mathrm{act}}\sim\chi n_s f_\tn{act}R/\gamma$, so that $\mathrm{Ca}_{\mathrm{act}}\sim1$ gives $R_c\sim\gamma/(\chi n_s f_\tn{act})$. Using $\gamma\sim5\,\mu\mathrm{N/m}$, $n_s\sim0.5\,\mu\mathrm{m}^{-2}$, $f_\tn{act}\sim1\,\mathrm{pN}$, and $\chi\sim1$ yields $R_c$ of order $10\,\mu\mathrm{m}$, comparable to the observed scale. This order-of-magnitude agreement identifies a crossover at which capillary bridge growth and active interfacial shear become comparable; it does not by itself determine the chain topology, which additionally reflects the geometry and wetting constraints of the intermediate-composition regime. Within this scaling picture, active shear becomes increasingly important relative to capillarity for $R\gtrsim R_c$, whereas no such dynamic arrest is expected when motility is removed, consistent with our experimental observations.

\begin{figure}[t!]
\centering \includegraphics[width=8.5cm]{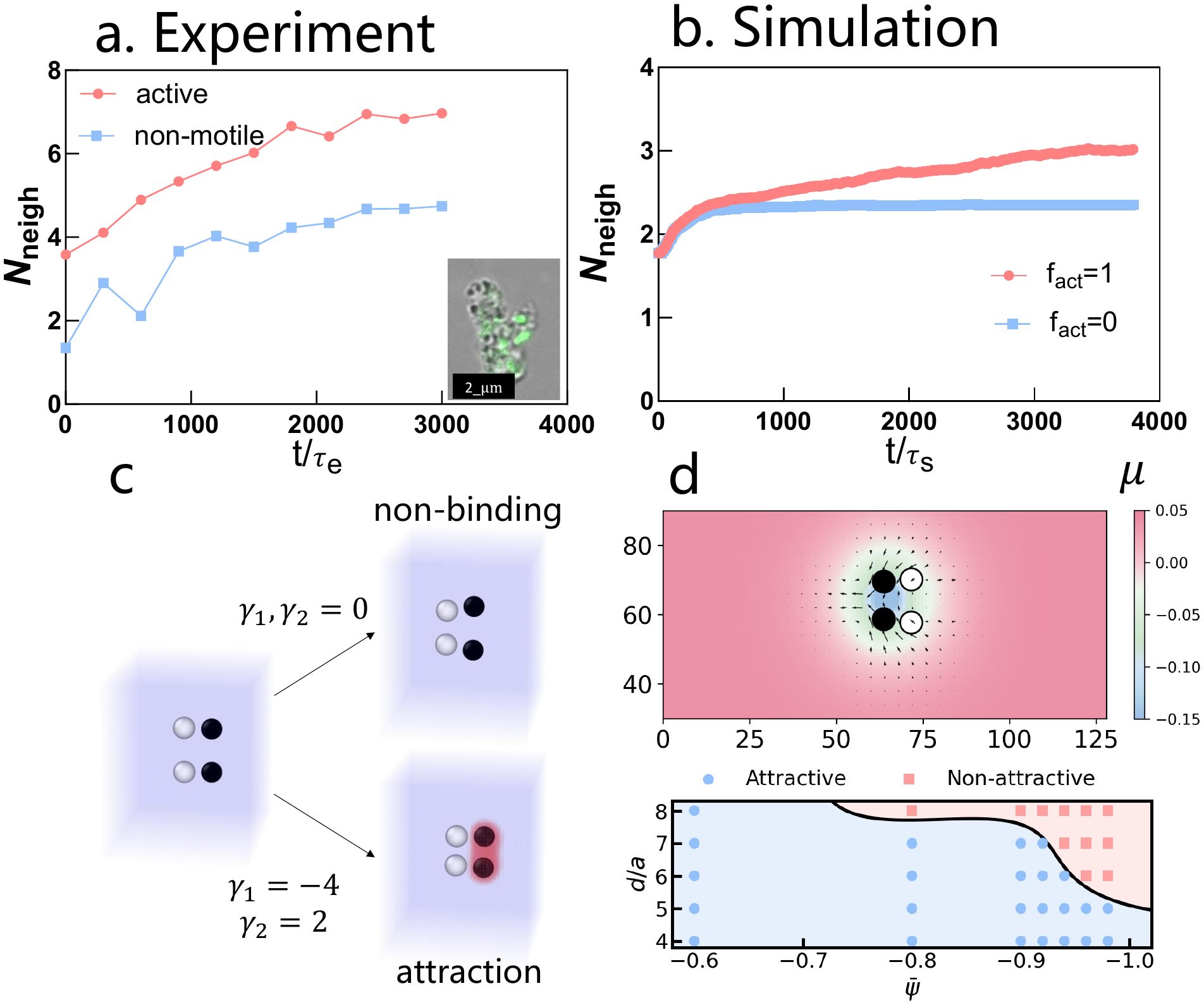}
\caption{Dynamics and wetting-mediated mechanism of branched bacterial clusters.
Time evolution of the mean local neighbors~$N_{\mathrm{neigh}}$ obtained from (a) experiments (Inset: a branched cluster) and (b) simulations. In both cases, active bacteria exhibit a faster increase and reach a higher steady-state $N_{\mathrm{neigh}}$ than passive controls, demonstrating enhanced bacterial aggregation kinetics in the wetting-bridged regime.
(c) Two-bacterium simulation showing that preferential wetting converts a non-binding interaction into a stable bound state ($\gamma_1=-4,\gamma_2=2$).
(d) Top: Chemical potential $\mu$ and velocity field $\bm{v}$, showing transient transport of the DEX-rich phase toward the inter-bacterial region.
Bottom: Binding states of two active bacteria versus initial separation $d/a_\tn{h}$ and mean composition $\bar{\psi}$.
}
\label{fig:figure3}
\end{figure}

\begin{figure}[t!]
\centering \includegraphics[width=8.5cm]{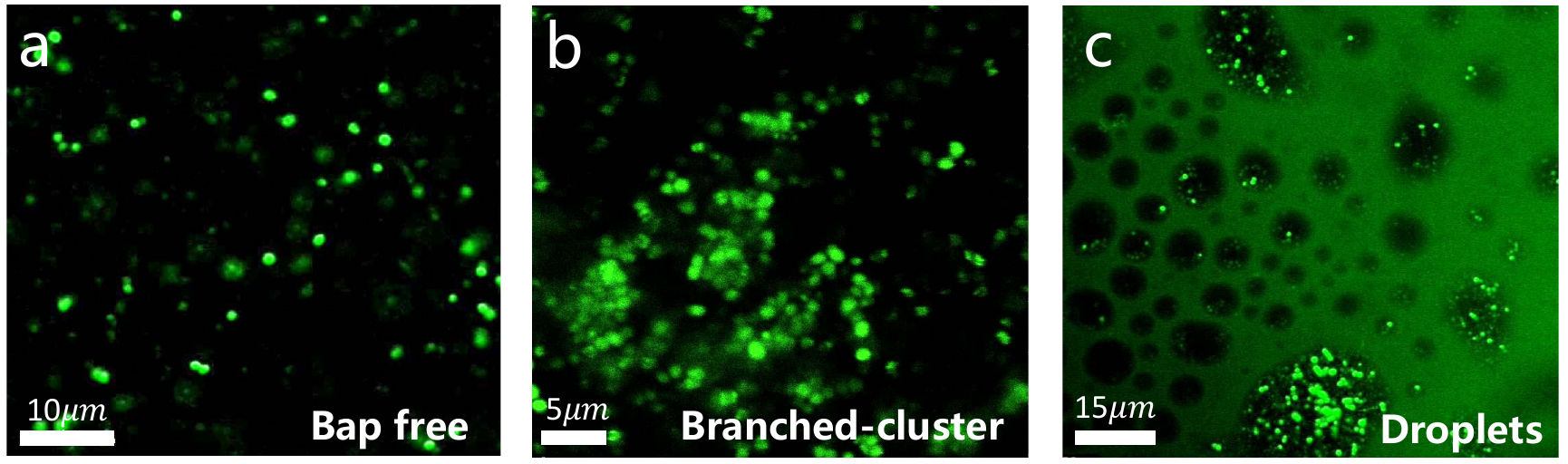}
\caption{Biofilm-associated protein (Bap) induces bacterial aggregation. Confocal images of \textit{P. aeruginosa} ($\phi_{\mathrm{bac}}=0.01$) in 8 wt\% PEG solutions.
(a) Bap-free control showing a dispersed state.
(b) With 1 $\mu$M Bap, bacteria form branched wetting-bridged clusters.
(c) With 10 $\mu$M Bap, bacteria are encapsulated within protein-rich droplets.}
\label{fig:figure4}
\end{figure}

{\it \bf Wetting-dominated branched clustering.---} As the DEX fraction is reduced further ($\phi_{\text{dex}} < \phi_{\text{bac}}$), the balance shifts toward wetting-mediated attraction, leading continuously to the branched, wetting-bridged state. The scarce DEX-rich phase then forms localized wetting patches and bridges between neighboring bacteria (Fig.~\ref{fig:figure1}(c), Fig.~\ref{fig:figureA1}(c), and Fig.~S3 in \cite{Supple}), producing a branched wetting-bridged cluster rather than isolated droplets. In this regime, the role of activity is reversed relative to the droplet regime: the measured neighbor number $N_{\mathrm{neigh}}(t)$ rises faster and reaches a larger value for active bacteria in both experiments and simulations (Fig.~\ref{fig:figure3}(a)--(b)), demonstrating that motility accelerates bacterial aggregation. At low $\phi_{\text{dex}}$, this aggregation is accompanied by growth of the DEX-rich domains (Fig.~S4 in \cite{Supple}), although $N_{\mathrm{neigh}}$ directly quantifies bacterial clustering rather than domain coarsening. Comparison with bacteria-free controls is not meaningful in this regime because the relevant structures are defined by the bacteria themselves.

The two-bacterium simulations clarify how preferential wetting enables binding ($\bar{\psi}=-0.98$; Fig.~\ref{fig:figure3}(c)--(d)). Without preferential wetting, the pair does not form a persistent bound state. When affinity for DEX is introduced, DEX-rich material accumulates between the bacteria and forms a stable wetting bridge. During their approach, chemical-potential gradients drive composition transport (Fig.~\ref{fig:figure3}(d)), accompanied by flow toward the inter-bacterial region, thereby facilitating bridge formation. These gradients are transient: as the bacteria approach, the chemical-potential gradient progressively weakens (Fig.~S5 in \cite{Supple}), while the pair remains bound after the gradient has largely relaxed. This persistence is consistent with thermodynamic stabilization by preferential wetting. Thus, nonequilibrium composition transport and the associated flow promote bacterial capture and bridge formation, whereas preferential wetting stabilizes the resulting bound state after the transient driving gradient has relaxed. Notably, the resulting capture can occur from micrometre-scale initial separations ($\approx 2\ell_\tn{bac}$; Fig.~\ref{fig:figure3}(d), bottom), providing a route to bacterial aggregation through a scarce wetting phase under dilute conditions~\cite{melaugh2016shaping,armbruster2018new,black2025capillary}.

{\it \bf Biological relevance of the wetting-bridged regime.---} To test whether wetting-assisted capture extends beyond the DEX--PEG system, we performed an \textit{in vitro} reconstitution using purified biofilm-associated protein that acts as an independent wetting component (Bap; Fig.~\ref{fig:figureA1}(a) and Fig.~S6 in \cite{Supple}). The bacteria were suspended in 8~wt\% PEG with no DEX present, so Bap provides an independent wetting component. At 1~$\mu$M Bap, the bacteria-free Bap/PEG solution remains macroscopically single phase (Fig.~S7(a) in \cite{Supple}), yet bacteria form branched clusters (Fig.~\ref{fig:figure4}(a)--(b)). At 10~$\mu$M, Bap undergoes bulk LLPS and forms protein-rich droplets that encapsulate the bacteria (Fig.~\ref{fig:figure4}(c) and Fig.~S7(b) in \cite{Supple}). Thus, Bap can promote bacterial assembly before visible bulk protein LLPS, supporting the broader concept of a biologically derived ``wetting glue.''

\noindent
{\it \bf Summary.---}We combined experiments and hydrodynamic simulations to reveal a composition-controlled crossover in active liquid--liquid phase separation. At high DEX fraction, interfacial motility generates rotational flow that disrupts liquid-bridge formation and suppresses coalescence. At intermediate composition, competition between this active shear and wetting-driven bridge formation is consistent with the emergence of long-lived connected chains. At low DEX fraction, wetting-mediated bridging dominates, while activity accelerates bacterial capture and aggregation. The persistence of bound pairs after chemical-potential gradients relax, together with the independent Bap experiments, identifies preferential wetting as the stabilizing thermodynamic ingredient. Together with our companion paper~\cite{yang2026b}, these results establish active interfacial flow versus wetting-driven bridge formation as a physical control principle for morphology and kinetics in active LLPS.

\noindent
{\it \bf Acknowledgments.}---
We thank Huihui Liu for performing part of the experiments and Tianyi Zhang for contributing to the partial implementation of the code.  J.Y. is supported by the National Natural Science Foundation of China (Grant No. 22503076) and the Guangdong Provincial Pearl River Talents Program (Grant No. 2024GA11X007). C.W. is supported by the Science and Technology Development Fund, Macao SAR (FDCT, Nos. 0010/2026/AIJ, 0039/2025/RIB1, 0024/2023/AFJ, and 0002/2025/NRP), the Guangdong Provincial Committee for Basic and Applied Basic Research (EF2023-00209-ICMS), as well as the Zhuhai UM Science \& Technology Research Institute (CP-102-2024). H.T. acknowledges support from the Grant-in-Aid for Specially Promoted Research (KAKENHI Grant No. JP20H05619) and Scientific Research (A) (KAKENHI Grant No. JP26H02008) from the Japan Society for the Promotion of Science (JSPS). J.Y. and D.Y. thank the HPC AI Intelligent Computing Platform at HKUST(GZ) for GPU resources.

\appendix
\centerline{\bf End Matter}
\renewcommand{\thefigure}{A\arabic{figure}}
\setcounter{figure}{0}   

\begin{figure}[b!]
\centering \includegraphics[width=8.5cm]{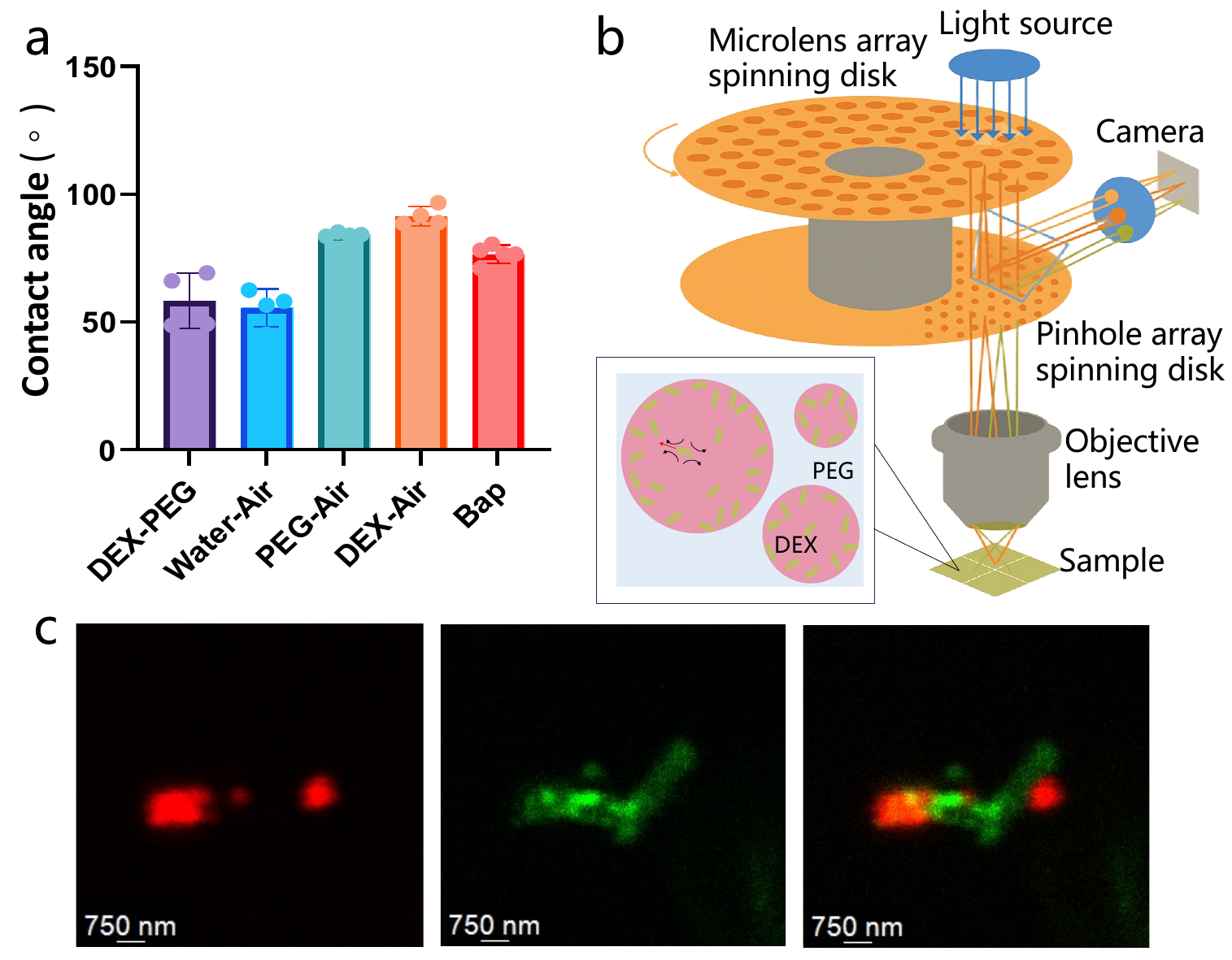}
\caption{Interfacial wettability and confocal imaging.
(a) Contact angles characterizing the wettability of \textit{P. aeruginosa} at the DEX--PEG and water/PEG/DEX--air interfaces and contact angles of Bap-rich phase on \textit{P. aeruginosa.}
(b) Schematic of the spinning-disk confocal microscopy setup for imaging the DEX--PEG system with \textit{P. aeruginosa}. (c) Direct evidence of DEX wetting around individual bacteria ($\phi_{\text{bac}} = 0.01$ and $\phi_{\text{dex}} = 0.001$). The panels display, from left to right: (i) the DEX phase (red), (ii) fluorescently labeled bacteria (green), and (iii) the merged overlay.
}
\label{fig:figureA1}
\end{figure}

\begin{figure}[t!]
\centering \includegraphics[width=8cm]{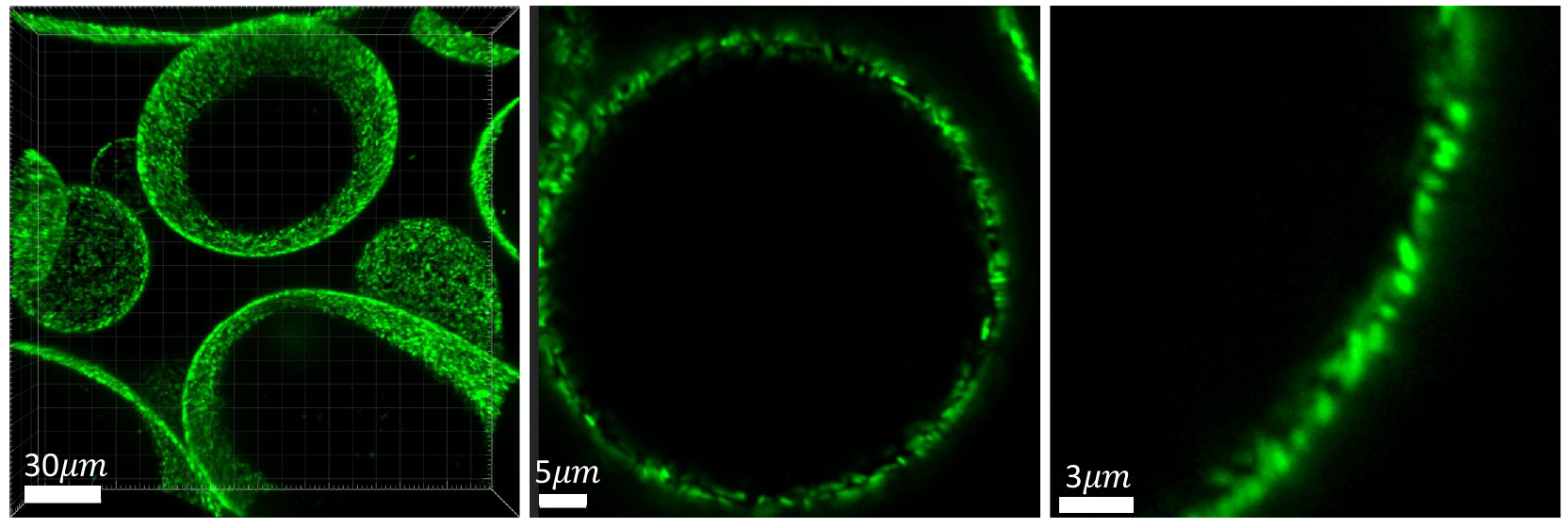}
\caption{Experimental evidence of \textit{P. aeruginosa} interfacial accumulation. Confocal microscopy images of fluorescently labeled \textit{P. aeruginosa} in DEX--PEG binary mixture ($\phi_{\mathrm{bac}}=0.01$ and $\phi_{\mathrm{dex}}=0.1$). This provides experimental evidence of interfacial accumulation and confirms the amphiphilic nature of \textit{P. aeruginosa}, supporting the physical basis of our model.
}
\label{fig:figureA2}
\end{figure}

\begin{figure}[t!]
\centering \includegraphics[width=8cm]{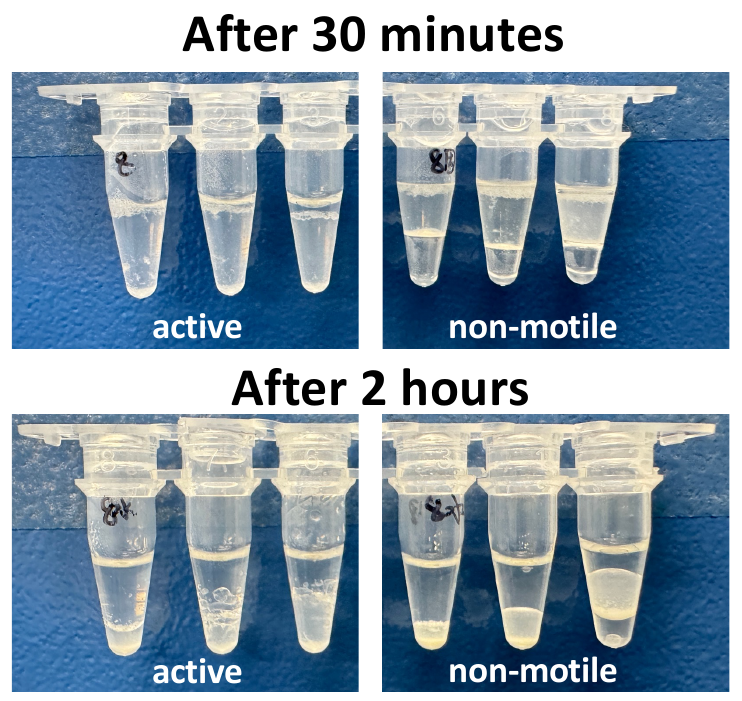}
\caption{Macroscopic phase separation in ATPS containing active and non-motile \textit{P. aeruginosa} at $\phi_{\mathrm{bac}}=0.05$. 
The DEX volume fraction $\phi_{\text{dex}}$ in each panel increases from left to right ($0.08$, $0.16$, $0.32$). 
After 30 minutes, non-motile systems exhibit complete macroscopic stratification, whereas active systems remain stable for up to 2 hours. 
}
\label{fig:figureA3}
\end{figure}

\begin{figure}[t!]
\centering \includegraphics[width=8cm]{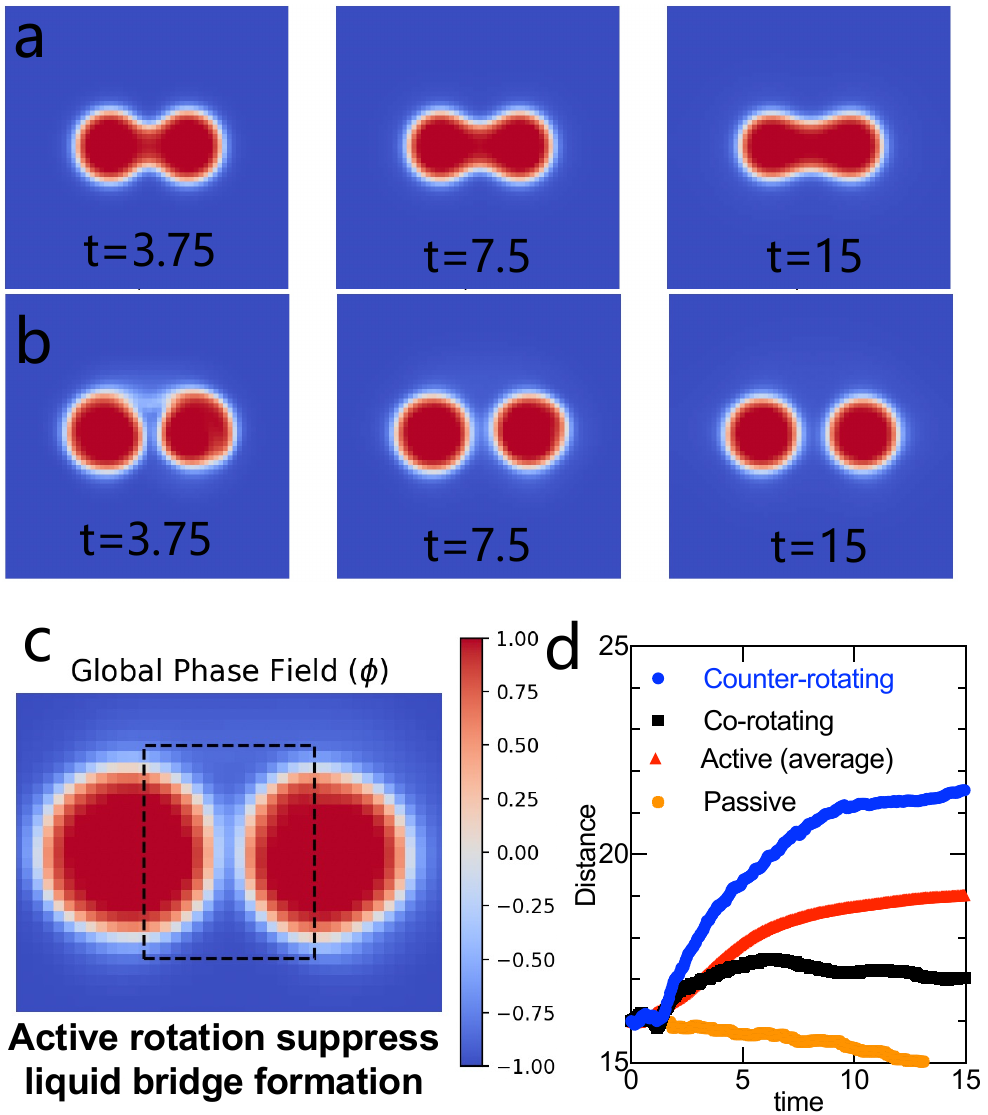}
\caption{Mechanistic validation of arrested coalescence by active interfacial flows. (a) Time-lapse snapshots of the coalescence process for inactive droplets without rotation. (b) Time evolution of active droplets with interfacial rotation, demonstrating arrested coalescence. (c)  Phase-field~$\psi$ in the two-droplet system. The coalescence process involves the establishment of a liquid bridge connecting the two droplets. Active rotation suppresses liquid bridge formation, thereby inhibiting the coalescence of droplets. (d) Time evolution of the center-to-center distance between the two droplets. The orange line represents the passive control where the distance decreases over time, indicating spontaneous droplet coalescence. The blue and black lines denote the counter-rotating and co-rotating droplets, respectively. The red line illustrates the averaged evolution across 1000 randomly assigned rotational axes, statistically confirming the suppression of droplet coalescence by interfacial flows.}
\label{fig:figureA4}
\end{figure}

\textbf{Contact angle measurement}. Contact angles were measured in two configurations (Fig.~S1(a) in \cite{Supple}). The first is a sessile-drop control (bacteria--liquid--air) that characterizes the overall surface wettability of the cells. \textit{P. aeruginosa} (ATCC 10145) was cultured overnight in LB (Solarbio, L8290) at $37^\circ\text{C}$. The cells were harvested by centrifugation (5,000~$\times$~g, 15 min, $4^\circ\text{C}$) and washed three times with deionized (DI) water to remove residual media components. The final pellet was resuspended in a small volume of DI water to create a dense cellular suspension. A 100~$\mu$L aliquot of this dense suspension was uniformly deposited onto a glass coverslip. The slides were then dried in a desiccator at room temperature overnight to form a stable, homogeneous bacterial lawn. The wettability of the dried bacterial lawn was assessed using a goniometer/drop shape analyzer at ambient temperature. A 2~$\mu$L droplet of the test liquid (DI water, 5~wt\% PEG-LB solution, or 5~wt\% DEX-LB solution) was gently deposited onto the surface of the bacterial lawn. The contact angle ($\theta$) was measured at the three-phase (solid-liquid-air) contact line immediately after the drop stabilized (Fig.~S1(b) and S1(d) in \cite{Supple}). Measurements were repeated at least five times at different locations of the lawn for statistical accuracy. The second is the three-phase (captive-drop) configuration (bacteria--DEX--PEG), which yields the contact angle that predicts the partitioning preference of the bacteria between the DEX and PEG phases. For this measurement, the bacterial lawn was mounted in a transparent cuvette. The cuvette was filled with the continuous phase liquid (5~wt\% PEG-LB solution), fully immersing the bacterial lawn. Using an inverted needle (J-needle), a single droplet (2~$\mu$L) of the second immiscible phase (5~wt\% DEX-LB solution) was carefully dispensed and placed onto the submerged bacterial lawn (``captive drop" configuration). The system was allowed to equilibrate for 5 minutes. The three-phase contact angle was then measured through the DEX-phase droplet (Fig.~S1(a) and S1(c) in \cite{Supple}). This angle represents the partitioning preference of the bacteria between DEX and PEG.

\textbf{Bap-mediated bacterial aggregation assay}. To investigate the role of Bap in mediating bacterial clustering and phase behavior, an \textit{in vitro} reconstitution assay was performed. \textit{Pseudomonas aeruginosa} strains expressing GFP were grown in LB medium at 37$^{\circ}$C to the exponential phase (OD$_{600} \approx 0.6$). The cells were harvested by centrifugation at 3,000~$\times g$ for 5~min and washed three times with PBS to remove culture medium and secreted factors. The bacterial pellet was resuspended in the same buffer, and the bacteria volume fraction ($\phi_{\text{bac}}$) was adjusted to approximately 0.01 for all experiments.
Aggregation mixtures were prepared by combining washed bacteria with purified recombinant Bap and an 8~wt\% PEG (MW 8000) crowding agent. To explore distinct aggregation regimes, the Bap concentration was tuned: 1~$\mu$M induced wetting-bridged branched aggregation, whereas 10~$\mu$M yielded protein-rich droplet structures.

\textbf{Hybrid model of bacteria in a binary mixture}. We propose a particle-field hybrid model for simulating the active bacteria immersed in a binary fluid mixture undergoing demixing, taking into account phase separation dynamics, hydrodynamic interactions (HIs), bacterial activity, and wetting effects. This simulation framework is broadly applicable to diverse motile bacteria and synthetic active particles in ATPS, where phase separation, self-propulsion, HIs, and wetting are intrinsically coupled. We employ a three-particle model to represent a single bacterium~\cite{furukawa2014activity} and incorporate HIs by directly solving the Navier--Stokes (NS) equation within the fluid particle dynamics (FPD) framework~\cite{tanaka2000,Furukawa2018}. Each particle is modeled as a viscous fluid particle with a smooth interfacial profile, and by choosing a high particle-to-solvent viscosity ratio ($\eta_{\rm p}=50\eta_{\rm s}$), the flow inside the particles is effectively suppressed so that each particle can be approximated as a rigid body; a complete free-energy functional is constructed, comprising the bulk Ginzburg--Landau free energy, surface wetting energy, particle impermeability constraint, and direct particle interactions, in which the amphiphilic nature of \textit{P.~aeruginosa} is captured through differential surface coupling strengths for the head ($\gamma_1=-4$) and tail ($\gamma_2=2$) particles, driving the bacterium to localize at the interface. The force density term in the NS equation is formulated to guarantee momentum conservation, coupling composition-field dynamics with the incompressible flow solved by the Marker-and-Cell (MAC) method on a staggered grid, thereby achieving a fully self-consistent simulation of the bacterium--binary-fluid--hydrodynamics coupling. The model equations and numerical implementation are described in detail in our companion study~\cite{yang2026b}.

To examine the role of rotational interfacial flow in droplet coalescence, we additionally simulate a pair of deformable phase-separated droplets using Stokes dynamics, comparing otherwise identical cases with and without imposed interfacial rotation (Fig.~\ref{fig:figureA4}). We monitor the phase field $\psi$, chemical potential $\mu$, local velocity $\bm{v}$, and the droplet center-to-center distance. For the active case in Fig.~\ref{fig:figure2}(e), the distance evolution is averaged over 1000 uniformly sampled random orientations of the rotational forcing.

\bibliography{journals,bac}

\end{document}